\documentclass[fleqn,12pt,twoside,a4paper]{article}
\usepackage{amsmath}
\usepackage{amssymb}
\usepackage{graphicx}
\usepackage{psfrag}
\usepackage{version}
\usepackage{dcolumn}

{
\begin{list}{}{\setlength{\rightmargin}{\leftmargin}\setlength{\topsep}{2pt}}
\item[]\ignorespaces\small\sf}
{\end{list}}

\newcommand{\op}[1]{\mathbf{#1}}

\newcommand{\Dmu}{\Delta\mu}

\newcommand{\zp}{Z_{+}}

\newcommand{\zm}{Z_{-}}
\newcommand{\V}[1]{\op{#1}}

\newcommand{\av}[1]{\langle#1\rangle}

\newcommand{\pp}{\widetilde{p}}

\newcommand{\ZG}[3]{Z_{#2,#3}(#1)}
\newcommand{\ZGu}[4]{Z_{#2,#3}^{#4}(#1)}

\newcommand{\p}[2]{p_{#1,#2}}
\newcommand{\ps}[2]{g_{#1;#2}}

\newcommand{\f}[2]{f_{#1,#2}}
\newcommand{\mt}{\widetilde{M}}
\newcommand{\Tt}{\widetilde{T}}

\newcommand{\lk}{\lambda^{(k)}}
\newcommand{\lo}{\lambda^{(0)}}     
\newcommand{\ak}{a^{(k)}}
\newcommand{\ao}{a^{(0)}}     
\newcommand{\Ak}{A^{(k)}}
\newcommand{\Ao}{A^{(0)}}     
\newcommand{\Bk}{B^{(k)}}
     
\newcommand{\uk}{\op{1}^{(k)}}
\newcommand{\uo}{\op{1}^{(0)}}

\begin{document}
\begin{flushright}
BP-TP 06/99
\end{flushright}
\begin{center}	\Large
{\bf Matter Correlations in Branched Polymers}\\[5mm]\large
Piotr Bialas$^{ab}$\\[2mm]
\small ${}^a$Fakult\"at fur Physik, Universit\"at Bielefeld, 33615 Bielefeld, Germany\\
\small ${}^b$Institute of Comp.~Science, Jagellonian University. 30-072 Krakow, Poland
\end{center}
\begin{abstract}
We analyze correlation functions in a toy model of a random geometry
interacting with matter.  We show that in general the connected
correlator will contain a long--range scaling part. This result
supports the previously conjectured general form of correlation
functions on random geometries.  We discuss the interplay between
matter and geometry and the role of the symmetry in the matter sector.
\end{abstract}

\section*{Introduction}

In theories with fluctuating (quantized) geometry it is non--trivial
to define the concept of a connected correlation function.  This
is especially apparent in the approaches like dynamical triangulation (DT)
which are formulated in a non--perturbative, coordinate--free way and
permit to go beyond the flat background expansion.  In such
formulations the distance itself is a dynamical and a highly non-local
object.  This leads to conceptual and technical difficulties when
trying to generalize the standard fixed--geometry definitions
\cite{deBakker:1995he,Bialas:1997ei}.

Due to the non--locality the correlation functions are very difficult
to study and very few analytic results exist
\cite{Ambjorn:1998vd,Kawai:1993cj}.  A growing body of numerical
evidence suggests however the existence of a simple structure common
to all correlators in various ensembles of random geometries
\cite{Ambjorn:1998vd}. This structure can be  studied
with toy models.

In this contribution we analyze a simple model of random geometry
interacting with matter which permits the detailed analytical analysis
of the connected and disconnected correlators.

The structure of this paper is as follows~: we first introduce the
model, then we derive its thermodynamical properties and general
structure of the correlation functions in grand--canonical and
canonical ensembles.  As an application of the developed formalism we
then investigate the case of Ising spins in magnetic field. We compare
our results with the Monte--Carlo simulations.

\section*{Branched Polymers}

Branched Polymers (BP) provide a very simple but non--trivial example
of a random geometry ensemble \cite{Ambjorn:1986dn}. Moreover they
exhibit a wide range of properties in common with higher dimensional
DT systems \cite{Bialas:1996eh}.

By Branched Polymers we understand the ensemble of {\em planar}, {\em
labeled} trees ($\mathcal{T}$) \cite{Ambjorn:1986dn,Harary:1973} .
Each tree $T$ is weighted with a factor $\rho(T,X)$ depending on the
geometry and possibly on the additional ``matter'' fields $X$ living
on the vertices of the tree. Partition function of the model is thus
defined to be~:
\begin{equation}\label{grandpartition}		
Z(\mu)=\sum_{T\in \mathcal{T}}\sum_{X}e^{-\mu n}\rho(T,X)
\end{equation}
where $n=n(T)$ denotes the number of vertices in the given tree. 
Considering the most general form of two--point actions
for geometry and matter fields we take $\rho$ to be~:
\begin{equation}\label{rho}
\rho(T,X)=\prod_i p_{q_i,x_i}\prod_{<i,j>}\ps{q_i,x_i}{q_j,x_j}.
\end{equation}
Here $q_i$ and $x_i$ denote respectively the number of branches
emerging from the vertex $i$ and the value of the matter field(s) in
this vertex, 
$<\!\!i,j\!\!>$ denotes the set of all nearest neighbors pairs.
Non--negative coefficients $\p{q}{x}$ and $\ps{q,x}{q',x'}$ are
parameters of the model. The only restriction is that we assume
$\ps{q,x}{q',x'}$ to be symmetric {\em ie}
$\ps{q,x}{q',x'}=\ps{q',x'}{q,x}$. 
The canonical partition function $Z_n$ is related to
 (\ref{grandpartition}) through the discrete Laplace transform~:
\begin{equation}	
Z(\mu)=\sum_{n=1}^\infty e^{-\mu n}Z_n.
\end{equation}

All the properties of this ensemble can be obtained from the partition
function of the ensemble $\mathcal{T}_{pl}$ of {\em rooted}, {\em planted} trees {\em ie}
the trees with one point (root) marked with the condition that this
marked vertex has only one branch
\cite{Harary:1973,Jurkiewicz:1997yd}.  This vertex thus constitutes a
``handle'' by which planted trees can be glued together. It is
convenient to assign to this vertex a dummy number of branches
$q_0$ and matter field $x_0$.
In this way we obtain
a whole set of partition functions for the planted ensemble~:
\begin{equation}\label{Zplanted}
\ZG{\mu}{q_0}{x_0}=\sum_{T\in\mathcal{T}_{pl}}\sum_{X}
e^{-\mu n} \prod_{i=1}^n\p{q_i}{x_i}\prod_{<i,j>} \ps{q_i,x_i}{q_j,x_j}.
\end{equation}
Here $n$ denotes the number of vertices not including the root.  The
sum over all neighbors includes the root (denoted by index zero).

Critical behavior of the model is characterized by singularities
of the functions $\ZG{\mu}{q}{x}$.  To find them we first observe that
those functions fulfill a set of equations (see
figure~\ref{partition})\cite{Ambjorn:1986dn}~:
\begin{figure}
\begin{center}	
\psfrag{qx}{$q$, $x$}
\psfrag{xp}{$x'$}
\includegraphics[width=12cm]{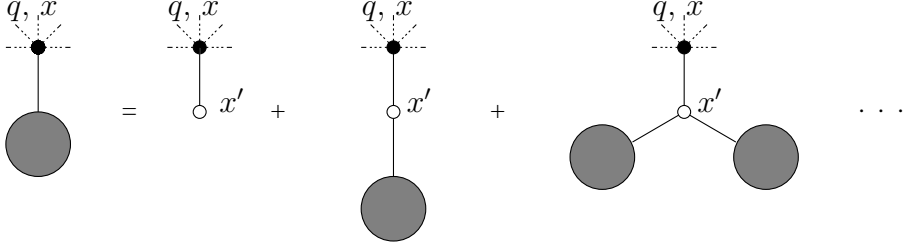}
\end{center}
\caption{\label{partition}Recursive construction of the partition
function $\ZG{\mu}{q}{x}$ of the ensemble of rooted planted trees.}
\end{figure}
\begin{equation}\label{sys}	
\ZG{\mu}{q}{x}=e^{-\mu}\sum_{q',x'}\, \ps{q,x}{q',x'}\,\p{q'}{x'}\,
\ZGu{\mu}{q'}{x'}{q'-1}.
\end{equation}
Expanding $\ZG{\mu}{q}{x}$ around  $\mu$~:
\begin{equation}\label{exp}	
\ZG{\mu+\Dmu}{q}{x}\approx \ZG{\mu}{q}{x}+\partial_\mu\ZG{\mu}{q}{x}\Dmu
\end{equation}
and inserting into (\ref{sys}) we obtain
\begin{equation}\label{lsys}
\sum_{q',x'}\bigl(\delta_{q,x;q'x'}-M_{q,x;q',x'}(\mu)\bigr)
\ZGu{\mu}{q'}{x'}{'}=-\ZG{\mu}{q}{x}
\end{equation}
where operator $\op{M}$ is defined by~:
\begin{equation}\label{M}	
M_{q,x;q',x'}(\mu)=e^{-\mu} \ps{q,x}{q',x'} \p{q'}{x'} 
(q'-1)\ZGu{\mu}{q'}{x'}{q'-2}. 
\end{equation}
Linear system (\ref{lsys}) has an unique solution provided that
the operator $\op{1}-\op{M}(\mu)$ is invertible.  The condition 
\begin{equation}\label{critcond}	
\op{1}-\op{M}(\mu_0)\quad \text{is {\em not} invertible} 
\end{equation}
defines the critical point $\mu_0$ and is equivalent to the statement
that operator $\op{M}(\mu_0)$ has an eigenvalue $\lo=1$.  As we
will see later, this corresponds to an {\em elongated} or {\em tree}
phase\cite{Bialas:1996ya}.

Around point $\mu_0$ expansion (\ref{exp}) is not valid.  Going beyond
the linear approximation, the first non--vanishing term will be in
general a quadratic form\footnote{Allowing negative values for $p$'s
it is possible to make also higher term of the expansion  vanish and by this 
 tune  the system to the multicritical point where
the corrections depend on $\Dmu^{\frac{1}{k}}$ \cite{Ambjorn:1990wp}.}
leading to~:
\begin{equation}\label{Zexpansion}	
\ZG{\mu_0+\Dmu}{q}{x}\approx 
\ZG{\mu_0}{q}{x}-\Delta\ZG{\mu_0}{q}{x}\sqrt{\Dmu}.
\end{equation}
As we show in the appendix~\ref{app:crit} $\Delta\ZG{\mu_0}{q}{x}$ is
 the eigenvector associated with the eigenvalue $\lo=1$.
We have chosen the sign in (\ref{Zexpansion}) as to make the
$\Delta\ZG{\mu_0}{q}{x}$ positive. This is ensured by the fact that
$\ZG{\mu}{q}{x}$ decreases as $\mu$ is increased as seen from 
(\ref{Zplanted}).

Condition (\ref{critcond}) is not the only way the singularity of the
partition function can arise.  The function of $\ZG{\mu}{q}{x}$ defined
by possibly infinite series on the right--hand side of (\ref{sys}) can
develop singularities.  This singularities will depend on the asymptotic
behavior of  $p$'s.  This scenario corresponds to {\em crumpled}
or {\em bush} phase. When both conditions are met simultaneously {\em
ie} $\lo=1$ and right--hand side of (\ref{sys}) is singular we may have
a {\em marginal} phase \cite{Jurkiewicz:1997yd,Bialas:1996ya}.

\begin{figure}[t]
\begin{center}
\psfrag{o}[c][c]{$\op{A}$}
\psfrag{o1u}{$A_{1,x}$}
\psfrag{o2u}[c][c]{$A_{2,x}$}
\psfrag{o3u}{$A_{3,x}$}
\psfrag{p}{$+$}
\psfrag{eq}{$=$}
\psfrag{cdots}[l][r]{$\cdots$}
\includegraphics[width=12cm,bb=25 0 594 123]{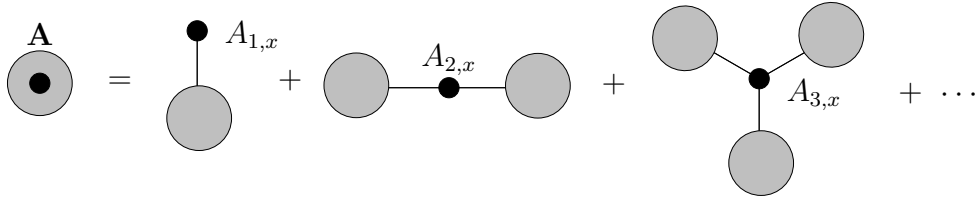}
\end{center}
\caption{\label{rooted}Rooted ensemble.}
\end{figure}

To calculate expectation values of a given operator $\op{A}$ we
introduce another kind of a partition function~:
\begin{equation}	
G^{\op{A}}(\mu)=
\sum_{T\in\mathcal{T}}\sum_{X}e^{-\mu n}\rho(T,X)
\sum_{i\in T} A_{q_i,x_i}
\end{equation}
which can be viewed as the partition function 
of the  ensemble of trees with one point (root) marked,
with operator $\op{A}$ inserted at the root~:
\begin{equation}	
G^{\op{A}}(\mu)=
\sum_{T\in\mathcal{T}_{root}}\sum_{X}e^{-\mu n}\rho(T,X)
A_{q_{0},x_{0}}.
\end{equation}
We can express $G^{\op{A}}(\mu)$ by the functions $\ZG{\mu}{q}{x}$
(see figure~\ref{rooted} and reference \cite{Harary:1973} for the
relation between planted and rooted ensembles)~:
\begin{equation}
G^\op{A}(\mu)=\sum_x  \sum_{q=1}^\infty 
A_{q,x} \frac{1}{q}\p{q}{x}\ZGu{\mu}{q}{x}{q}.
\end{equation}
Assuming that we are in the elongated phase
we can expand the above expression using  (\ref{Zexpansion}) to obtain~:
\begin{equation}\begin{split}	
G^{\op{A}}(\mu)&\approx\sum_x\sum_{q=1}^\infty 
{A}_{q,x} \p{q}{x}\ZGu{\mu_0}{q}{x}{q}\frac{1}{q}
\bigl(1-q\frac{\Delta\ZG{\mu_0}{q}{x}}{\ZG{\mu_0}{q}{x}}\sqrt{\Dmu}\bigr)\\
&\approx\sum_x \sum_{q=1}^\infty 
{A}_{q,x} \p{q}{x}\ZGu{\mu_0}{q}{x}{q}\frac{1}{q}\exp\bigl(-q\frac{\Delta\ZG{\mu_0}{q}{x}}{\ZG{\mu_0}{q}{x}}\sqrt{\Dmu}\bigr).
\end{split}
\end{equation}
Performing the inverse Laplace transform we obtain the  
large $n$ behavior of the canonical  function~: 
\begin{equation}\label{Un}	
G_n^\op{A}\approx\frac{C}{n^{\frac{3}{2}}}
e^{\mu_0 n}\sum_x\sum_{q=1}^\infty 
{A}_{q,x} \p{q}{x}\ZGu{\mu_0}{q}{x}{q-1}\Delta\ZG{\mu_0}{q}{x}.
\end{equation}
We get the expectation value of operator $\av{\op{A}}_n=\frac{1}{n}\av{\sum_i A_{q_i,x_i}}_n$ in canonical ensemble
by normalizing  the function $G_n^\op{A}$~:
\begin{equation}\label{avg}
\av{\op{A}}_n=
\frac{G_n^\op{A}}{G_n^\op{1}}\stackrel{N\rightarrow\infty}{\approx}
\frac{\sum_x\sum_{q=1}^\infty 
\op{A}_{q,x} \p{q}{x}\ZGu{\mu_0}{q}{x}{q-1}\Delta\ZG{\mu_0}{q}{x}}
{\sum_x\sum_{q=1}^\infty \p{q}{x}\ZGu{\mu_0}{q}{x}{q-1}\Delta\ZG{\mu_0}{q}{x}}.
\end{equation}

\section*{Correlation functions}

We define the integrated (unnormalized)  correlation functions
of operators $\op{A}$ and $\op{B}$  as~:
\begin{equation}\label{GAB}	
G^{\mathbf{AB}}_\mu(r)=\sum_{T\in\mathcal{T}}\sum_{X}\rho(T,X)\sum_{i,j \in T}
A_{q_i,x_i}B_{q_j,x_j}\delta_{d(i,j),r}.
\end{equation}
The distance $d(i,j)$ is the {\em geodesic} distance {\em ie} the
length of the shortest path between points $i$ and $j$. In our case
this definition is especially simple as only one path exists between
two points on a tree.
\begin{figure}[t]
\begin{center}
\psfrag{q0}{$\scriptstyle q_0,x_0$}	
\psfrag{q1}{$\scriptstyle q_1,x_1$}	
\psfrag{qi}{$\scriptstyle q_i,x_i$}	
\psfrag{qip}{$\scriptstyle q_{i+1},x_{i+1}$}
\psfrag{qr}{$\scriptstyle q_r,x_r$}		
\psfrag{A}[c][c]{$\scriptstyle A_{q_0,x_0}$}		
\psfrag{B}[c][c]{$\scriptstyle B_{q_r,x_r}$}		
\psfrag{T}[c][c]{$\underbrace{\phantom{xxxxxx}}%
_{\ps{q_i,x_i}{q_{i+1},x_{i+1}}(q_{i+1}-1)\ZGu{\mu}{q_{i+1}}{x}{q_{i+1}-2}}$}
\includegraphics[width=12cm]{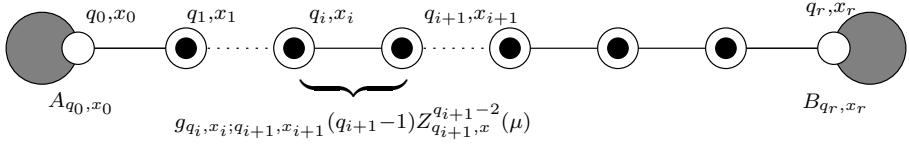}	
\end{center}
\caption{\label{correlation}Correlation function}
\end{figure}

To calculate the functions (\ref{GAB}) we consider the chain of $r+1$
points along the path joining points $i$ and $j$ and we sum
over all possible values of the coordination numbers and matter fields
values in each vertex (see figure~\ref{correlation}).  This can be
done by the transfer matrix technique.  In addition to one--point
weight each vertex inside the chain with coordination number $q$ has
$q-2$ trees attached to it. Those trees can be arranged in $q-1$ ways
relatively to the chain.  This gives an additional
$(q-1)\ZGu{\mu}{q}{x}{q-2}$ factor per
vertex\cite{Ambjorn:1990wp,Bialas:1996xq}.  If in constructing the
transition matrix we assign the one--point weights to the right--hand
side, we end up with a matrix that is {\em identical} to the operator
$\op{M}$ defined by (\ref{M}) except that the first row and
column are zero as vertices with order one cannot appear inside the
chain (see figure~\ref{correlation}).  We will call this truncated matrix $\op{\mt}$.

In terms of matrix $\op{\mt}$ the correlation function (\ref{GAB})
can be expressed as~:
\begin{multline}\label{GAB:2}	
G^{\op{AB}}_\mu(r) = \sum_{q,x}\sum_{q',x'}
\ZG{\op{A}}{q}{x}
\bigl(\op{M}\;
\op{\mt}^{r-2}\;
\op{M}\bigr)_{q,x;q'x'}
\frac{\ZG{\op{B}}{q'}{x'}}{(q'-1)\ZGu{\mu}{q'}{x}{q'-2}}
\end{multline}
with
\begin{equation}	
\ZG{\op{A}}{q}{x}=e^{-\frac{1}{2}\mu_0}A_{q,x}\, \p{q}{x}\, \ZGu{\mu}{q}{x}{q-1}.
\end{equation}
The functions $\ZG{\op{A}}{q}{x}$ account for the end--point effects
and the apparent asymmetry  is compensated by the 
asymmetry of matrix the $\op{\mt}$.
The above expression can be expanded in the basis of the 
eigenvectors of the matrix $\op{M}$ (see appendix~\ref{app:tm} for details)~: 
\begin{equation}\label{GABres}
G^{\op{AB}}_\mu(r)=\sum_k\bigl(\lk\bigr)^r\, \;\sum_{q,x}
\ZG{\op{A}}{q}{x}\,v^{(k)}_{q,x}\;
 \sum_{q',x'}\ZG{\op{B}}{q'}{x'}\, v^{(k)}_{q',x'}.
\end{equation}
The vectors $\V{v}^{(k)}$ are the right eigenvectors of the matrix
$\op{M}(\mu)$. They normalization is defined   through the relations
(\ref{u2v}) and (\ref{v2v}) in appendix~\ref{app:tm}.

The formula (\ref{GABres}) is exact and  valid for any
BP system with weights defined by (\ref{rho}).  We will now
expand this expression around $\mu_0$ assuming that we are in
elongated phase. From (\ref{Zexpansion}) we expect~:
\begin{equation}\begin{split}\label{lambdaexpansion}	
\lambda_k&\approx \lk_{0}-\lk_{1}\sqrt{\Dmu},\\
\sum_{q,x}\ZG{\op{A}}{q}{x}v^{(k)}_{q,x}&\approx \Ak_{0}-\Ak_{1}\sqrt{\Dmu}.
		\end{split}
\end{equation}

As we have shown in the previous section, in elongated phase,
$\lo_{0}=1$ and $v^{(0)}_{q,x} \propto \Delta\ZG{\mu_0}{q}{x}$.
Comparing (\ref{lambdaexpansion}) with (\ref{Un}) and (\ref{avg}) we
find that ${\Ao_{0}}=\uo_0 \av{\op{A}}$ (by $\op{1}$ we denote a
constant operator equal to unity : $1_{q,x}=1$).

Inserting (\ref{lambdaexpansion}) into (\ref{GABres}) and using
approximation $(1-x)^r\approx e^{-x \cdot r}$ we finally obtain the
structure of an arbitrary correlation function in the elongated phase
in the limit $\Dmu\rightarrow0$ with $\sqrt{\Dmu} \cdot r = const
\lesssim 1$~:
\begin{equation}\label{GABapprox}\begin{split}	
G^{\op{AB}}_\mu(r)&\approx
\bigl(\uo_0\bigr)^2\av{\op{A}}\av{\op{B}}
e^{-\ao( r + 
\delta^{(0)}_{\op{A}}+\delta^{(0)}_{\op{B}})\sqrt{\Dmu}}\\
&\phantom{==}+\sum_{k>0} 
\Ak_{0}\Bk_{0}(\lk_{0})^r
e^{-\ak( r +
\delta^{(k)}_{\op{A}}+\delta^{(k)}_{\op{B}})\sqrt{\Dmu}}
		\end{split}
\end{equation}
where
\begin{equation}\label{shifts}	
\ak=\frac{\lk_{1}}{\lk_{0}},\quad\text{and}\quad
\delta^{(k)}_{\op{A}}=\frac{1}{\ak}\frac{\Ak_{1}}{\Ak_{0}}
\end{equation}

At this stage we want to make two remarks~:
i) The first term in (\ref{GABapprox}) is a function of only one variable
$x=(r+\delta^{(0)}_{\op{A}}+\delta^{(0)}_{\op{B}}) \sqrt{\Dmu}$ and so
the distance scale is set by $\frac{1}{\sqrt{\Dmu}}$.  As we approach
the critical value $\mu_0$ (thermodynamic limit) the size of the
system becomes infinite. This justifies the name elongated phase. In
this phase Hausdorff dimension $d_H=2$ (this will be more apparent in
the canonical formulation). In thermodynamic limit shifts
$\delta^{(0)}_{\op{A}(\op{B})}$ can be neglected but they are
necessary to maintain the scaling at any finite volume. This fact was
first observed in MC simulations of 2D random surfaces
\cite{Ambjorn:1995rg}.

ii) If there exists a finite gap between $\lo_0=1$ and the
``next'' eigenvalue $\lambda =\limsup_{k>0}\,\lk_0$
then the first term will
dominate at large distances~:
\begin{equation}\label{GABassympt}	
G^{\op{AB}}(r)\approx \av{\op{A}}\av{\op{B}}
G^{\op{11}}\bigl(r+\delta^{eff.}_{\op{A}} +\delta^{eff.}_{\op{B}} \bigr)
\quad \text{for}\quad r \gg \xi\equiv-\frac{1}{\log \lambda}
\end{equation}
with
\begin{equation}	
\delta^{eff.}_{\op{A}}=\delta^{(0)}_\op{A}-\delta^{(0)}_\op{1}.
\end{equation}
The function $G^{\op{11}}(r)\equiv G(r)$ contains no operator insertions. 
Properly  normalized it gives average volume of the spherical shell of radius
$r$. As such it encodes the general information about the average shape and 
size of the system. The formula (\ref{GABassympt}) states that the whole 
effect of inserting operators is limited to  multiplicative factors 
and shifts of the argument. The shifts are additive which means they depend 
on each operator separately.

Obviously the formulas (\ref{GABapprox}) and (\ref{shifts}) are valid
only when $\Ak_0$, $\lk_0$ and $\lk_1$ are non--zero.  We will assume
that this is the case for $\lambda$'s and discuss the case of
$\Ak_0=0$.  Expanding (\ref{lambdaexpansion}) to the next order~:
\begin{equation}	
\sum_{q,x}\ZG{A}{q}{x}v^{(k)}_{q,x}\approx -\Ak_1\sqrt{\Dmu}+\Ak_2 \Dmu
\end{equation}
we find the corresponding term in the sum (\ref{GABapprox}) to be~:
\begin{multline}\label{zeroav}	
-\Ak_1\sqrt{\Dmu}\;
\exp\biggl({\displaystyle-\ak\bigl(r+\frac{\Ak_2}{\Ak_1}\frac{1}{\ak}\bigr)\sqrt{\Dmu}}\biggr)
=\\
{\Ak_1}\frac{1}{\ak}\frac{\partial }{\partial r}\;
\exp\biggl({\displaystyle-\ak\bigl(r+\frac{\Ak_2}{\Ak_1}\frac{1}{\ak}\bigr)\sqrt{\Dmu}}\biggr). 
\end{multline}
We will use this result in the next section when discussing the
connected correlators.

\subsection*{Connected correlators}

To study the connected correlators we switch now to the canonical
ensemble.  Connected correlation functions $G^{\op{AB}}_n(r)$ are
defined in the same way as grand--canonical ones, but now summation in
(\ref{GAB}) is restricted to the canonical ensemble. 
Additionally we now normalize correlation functions so that
$\sum_r G_n^{\op{11}}(r)\propto n$.  Taking the inverse Laplace
transform off (\ref{GABapprox}) we obtain~:
\begin{equation}\label{GABnapprox}\begin{split}	
G^{\op{AB}}_n(r)&\approx
2\sqrt{ n}\bigl(\uo_{0}\bigr)^2\av{\op{A}}\av{\op{B}}
g\bigl(\ao\frac{r + 
\delta^{(0)}_{\op{A}}+\delta^{(0)}_{\op{B}}}{2\sqrt{ n}}\bigr)\\
&\phantom{==}+2\sqrt{ n}\sum_{k=1} 
\Ak_0\Bk_0\bigl(\lk_0\bigr)^r
g\bigl(\ak\frac{r + 
\delta^{(k)}_{\op{A}}+\delta^{(k)}_{\op{B}}}{2\sqrt{n}}\bigr)
		\end{split}
\end{equation}
where $g(x)=x e^{-x^2}$. 
The structure of the correlation functions is thus the same as for 
the grand--canonical ensemble with $n$ playing the role of $1/\Dmu$
and $g(x)$ substituted for $e^{-x}$. In particular the asymptotic form
(\ref{GABassympt}) is also valid in this ensemble.   

We define the connected correlator by\footnote{See \cite{deBakker:1995he} 
and \cite{Bialas:1997ei} for discussion
of other possible definitions.}~:
\begin{equation}\label{conndef}\begin{split}
G^{\op{AB}conn}&=G^{\op{A}^{conn}\op{B}^{conn}}_n(r)
=G^{\op{(A-\av{\op{A}})(B-\av{\op{B}})}}_n(r)\\
&=G^{\op{AB}}_n(r)-\av{\op{A}}G_n^{\op{1B}}(r)-\av{\op{B}}G^{\op{1A}}_n(r) + \av{\op{A}}\av{\op{B}}G^{\op{11}}_n(r).
				\end{split}
\end{equation}
The operator $A^{con}_{q,x}=A_{q,x}-\av{\op{A}}$  has a zero average.
It is easy to check that~:
\begin{equation}\label{Aconn}
A^{(1)conn}_1=\ao\op{1}^{(0)}_0\av{\op{A}}\delta_{\op{A}}^{eff}\quad
\text{and}\quad\begin{aligned}
A^{(k)conn}_0=A^{(k)}_0-\av{\op{A}}\op{1}^{(k)}_0\\
A^{(k)conn}_1=A^{(k)}_1-\av{\op{A}}\op{1}^{(k)}_1\\
\end{aligned}
\end{equation}
and  thence from formula (\ref{zeroav}) we obtain~:
\begin{multline}\label{GABnconn}	
G^{\op{AB}conn}_n(r)\approx
\frac{\delta^{eff.}_{\op{A}}\delta^{eff.}_{\op{B}}}{2\sqrt{ n}}
\op{1}_{0,0}^2\av{\op{A}}\av{\op{B}}a^2_{0}\;
g''\bigl(\ao\frac{r+\cdots}{2\sqrt{n}}\bigr)\\
+2 \sqrt{n}\sum_{k=1}
\op{1}_{k,0}^2
\biggl(\frac{\Ak_0}{\uk_0}-\av{\op{A}}\biggr)
\biggl(\frac{\Bk_0}{\uk_0}-\av{\op{B}}\biggr)
\bigl(\lk\bigr)^r 
\, g\bigl(\ak\frac{r+\bar\delta^{(k)}_{\op{A}}+\bar\delta^{(k)}_{\op{B}}}{2\sqrt{n}}\bigr).
\end{multline}
The dots stand for the shifts which depend on $A^{(0)}_2$ and
$B^{(0)}_2$ (see (\ref{zeroav})). The shifts  $\bar\delta^{(k)}_{\op{A(B)}}$
are obtained by inserting (\ref{Aconn}) into (\ref{shifts}). 

A more intuitive way of obtaining
this results is to insert (\ref{GABapprox}) into the expanded
expression for the connected correlator
(\ref{conndef})\cite{Ambjorn:1998vd,Bialas:1998jz}.

Comparing (\ref{GABnconn}) with (\ref{GABnapprox}) we see that the
scaling long--range part is suppressed by factor of $n$.  We will call
this weak long--range correlations\footnote{This is in analogy with
the model of spins interacting with infinite range forces but with
strength diminishing with the system size and leading to similar
behavior of the connected correlation functions $G^{conn}_n(r)\sim
\frac{f(r)}{n}$ \cite{Parisi:1988}.}.  The non--scaling terms remain
of the leading--order in $n$ and for any finite $r$ will eventually
dominate in the thermodynamic limit.  Hoverer when considered as
functions of the scaling variable $x=r/\sqrt{n}$ they will vanish in this
limit.

The weak long--range correlations and can be further suppressed by
using an ``improved'' version of the connected
correlator \cite{Ambjorn:1998vd}~:
\begin{multline}\label{GABimp}	
G^{\op{AB}conn}_n(r)=
              G^{\op{AB}}_n(r)
-\av{\op{A}}_nG^{\op{1B}}(r+\delta^{eff.}_\op{A})
-\av{\op{B}}G^{\op{1A}}_n(r+\delta^{eff.}_\op{B}) \\
+\av{\op{A}}\av{\op{B}}
G^{\op{11}}_n(r+\delta^{eff.}_\op{A}+\delta^{eff.}_\op{B}).
\end{multline}
Inserting this into (\ref{conndef}) and omitting the shifts we obtain~:
\begin{multline}	\label{GABimproved}
G^{\op{AB}conn}_n(r)\approx
\frac{1}{n}
C \; g'''\bigl(\ao \frac{r}{2\sqrt{n}}\bigr)\\
+2\sqrt{n}\sum_k
\op{1}_{k,0}^2
\biggl(\frac{\Ak_0}{\uk_0}-\av{\op{A}}\lambda_k^{\delta_\op{A}}\biggr)
\biggl(\frac{\Bk_0}{\uk_0}-\av{\op{B}}\lambda_k^{\delta_\op{B}}\biggr)
\bigl(\lk_0\bigr)^{r}
g\bigl(\ak \frac{r}{2\sqrt{n}}\bigr).
\end{multline}
We see that the scaling term is
further suppressed by a factor $\sqrt{n}$ while the non--scaling terms
remain of the leading--order.

\section*{Ising model}

In this section we consider the Ising model on BP as defined in
\cite{Ambjorn:1993rp}. This corresponds to the choice of
$\ps{q,s}{q's'}=e^{\beta s\cdot s'}$ and $\p{q}{s}=p_q\cdot e^{-h s}$.
Spins take values $\pm1$.  Because we have no geometric interactions,
partition function $\ZG{\mu}{q}{x}$ does not depend on $q$. Summing
the right--hand side of (\ref{sys}) over $q'$ we obtain
\cite{Ambjorn:1993rp}~:
\begin{equation}\label{zpm}
\begin{split}	
\zp(\mu)&=e^{-\mu}\biggl(e^{\beta}e^{h} F\bigl(\zp(\mu)\bigr)+e^{-\beta}e^{-h} F\bigl(\zm(\mu)\bigr)\biggr)\\
\zm(\mu)&=e^{-\mu}\biggl(e^{-\beta}e^{h} F\bigl(\zp(\mu)\bigr)+e^{\beta}e^{-h} F\bigl(\zm(\mu)\bigr)\biggr)
\end{split}
\end{equation}
where~:
\begin{equation}
F(Z)=\sum_{q=1}^\infty p(q)Z^{q-1}
\end{equation}
The critical point condition (\ref{critcond}) is now~:
\begin{multline}\label{sing}	
1=
e^{-\mu_0}e^{\beta}\biggl(e^hF'\bigl(\zp(\mu_0)\bigr)
+e^{-h}F'\bigl(\zm(\mu_0)\bigr)\biggr)\\
-e^{-2\mu_0} 2 \sinh(2\beta) 
F'\bigl(\zp(\mu_0)\bigr)F'\bigl(\zm(\mu_0)\bigr).
\end{multline}
The transfer matrix $\op{\op{M}}$ (there is no distinction
between $\op{\mt}$ and $\op{M}$ as we no longer have the dependency on
$q$) 
is~: 
\begin{eqnarray}\label{tran}	
\op{M}=e^{-\mu}
\begin{pmatrix}
e^\beta e^h F'(\zp) & e^{-\beta}F'(\zm) \\[3mm]
e^{-\beta}F'(\zp) &e^\beta e^{-h} F'(\zm)
\end{pmatrix}
\end{eqnarray}
From (\ref{GABnconn}) we have~: 
\begin{multline}\label{XY}	
G^{\op{AB}conn}_n(r)\approx 
\frac{1}{2 \sqrt{ n}}\, 
X_{\op{AB}}\,
g''\bigl(\ao \frac{r+\cdots}{2\sqrt{n}}\bigr)\\ + 
2\sqrt{ n}\,
Y_{\op{AB}}\,\bigl(\lambda^{(1)}_0\bigr)^r\, g\bigl(a^{(1)} \frac{r+\cdots}{2\sqrt n}\bigr)  
\end{multline}
We will refer to the first term of the above expression as to  ``scaling'' term and to the other
as to ``non--scaling'' one. 
The expressions for $X(Y)_{\op{AB}}$ can be readout from
(\ref{GABnconn}).  The case of the ``curvature--curvature'' 
correlator\footnote{In DT the curvature is related to the number of $D$--simplices 
incident on a $(D-2)$--simplex}
$G^{\op{qq}}_n(r)$ is special (operator $\op{q}$ is defined by
$q_{u,x}=u$). We can obtain it directly from $G^{\op{11}}_n(r)$ using
the relations~:
\begin{equation}\label{gtree1}\begin{split}		
G^{\op{Aq}}(r)&=G^{\op{A1}}(r+1)+G^{\op{A1}}(r),\quad r>0\\
G^{\op{Aq}}(0)&=G^{\op{A1}}(1)\quad\text{and}\quad
\av{\op{q}}=2\\
		\end{split}
\end{equation}
that are satisfied on any tree.
Inserting the above relations into
  (\ref{conndef}) we obtain~:
\begin{equation}\label{qqcon}\begin{split}		
G^{\op{qq}conn}_n(r)&=
G^{\op{11}}(r+2)-2G^{\op{11}}(r+1)+G^{\op{11}}(r)\approx G^{''\op{11}}(r+1)
		\end{split}
\end{equation}
 
When $h=0$  the symmetry implies $\zp(\mu)=\zm(\mu)$ and 
the equations  (\ref{sys})  decouple.  
The only effect of the presence of Ising spins is to renormalize
the chemical potential \cite{Ambjorn:1993rp}.  This leads to~:
\begin{equation}	
G^{\op{ss}}_n(r)=G^{\op{ss}conn}_n(r)=\bigl(\tanh\beta\bigr)^r \mathcal{G}^{\op{11}}_n(r)
\end{equation}
where 
\begin{equation}	
\mathcal{G}^{\op{11}}_n(r)\approx C \sqrt{n} r e^{-a\frac{r^2}{n}}
\end{equation}
is the point--point correlation function of the pure (without spins)
BP model.  The correlation functions of the operators
depending only on the geometry remain unaffected by the presence
of spins.  
We see  that in this special case matter and geometry
sectors do not interact~:
\begin{equation}	
X_{\op{ss}}=0 \quad \text{and}\quad Y_{\op{qq}}=0.
\end{equation}
The vanishing of $X_{\op{ss}}$ is actually a general feature~:
whenever symmetry enforces the condition $\ZG{\mu}{q}{x}= Z_{q}(\mu)$
the matter--matter correlators will not contain the long--range
scaling part~: $X_{xx}=0$ (see Appendix~\ref{app:symm} for the
proof).

When $h\neq 0$ the symmetry is explicitly broken and we have to
investigate the full  system (\ref{zpm},\ref{sing}).  
Those equations 
cannot be solved analytically even for the simplest choices of
$p$'s. They are however not difficult to solve numerically. Once
$\mu_0$ and $Z_\pm(\mu_0)$ are known one can calculate other necessary
quantities.  To obtain the $\sqrt{\Dmu}$ order terms the standard 
perturbation theory can be used.  

For simplicity we have chosen the weights $p_1=p_2=p_3=1$ and all
others equal to zero. The inverse temperature and magnetic field were
set to $\beta=1.0$ and $h=0.1$.  For those values we obtained
$\lambda^{(1)}_0\approx0.5709$ corresponding to spin--spin
correlation length $\zeta\approx 1.785$. For $\ao$ and and $a^{(1)}$
we obtained respectively $1.1795$ and $0.5760$.  We have compared those
theoretical large $n$ limit predictions with the results of MC
simulations of the systems with 250, 1000 and 4000 vertices.

\begin{table}[t]
\begin{center}	
\begin{tabular}{||r||D{.}{.}{7}|D{.}{.}{6}|c|}
\hline\hline
 $n$ 	& \multicolumn{1}{c|}{$\av{\op{s}}$}&  
\multicolumn{1}{c|}{$\delta^{eff.}_\op{s}$} & \multicolumn{1}{c|}{$E^2$} \\
 {\it range} 	& 	&  & \\\hline\hline
{\it theor.}& 0.6980 & -0.6533 &          \\\hline\hline
250 	& 0.6961(3)  & -0.647(4)  & 0.20e-3   \\\cline{2-4}
10--80        & 0.6962(3)  & -0.625(4)  & 0.90e-4   \\\hline\hline
1000 	& 0.6979(1)  & -0.644(1)  & 0.10e-2   \\\cline{2-4}
10--160        & 0.6979(1)  & -0.652(1)  & 0.15e-3  \\\hline\hline
4000	& 0.69798(7) &  -0.649(1) & 0.56e-4  \\\cline{2-4}
10--200        & 0.69799(7) & -0.652(1)  & 0.26e-3  \\\hline\hline
\end{tabular}
\end{center}
\caption{\label{shift}Fitted values of shifts and average spin values.}
\end{table}
First we have tested the validity of the approximation
(\ref{GABassympt}). 
For  \mbox{$r\gg \zeta$} we expect 
\begin{equation}\label{fitform}	
G^{\op{1s}}(r)\approx \av{\op{s}}G(r+\delta^{eff.}_{\op{s}})
\quad\text{and}\quad 
G^{\op{ss}}(r)\approx \av{\op{s}}G^{\op{s}}(r+\delta^{eff.}_\op{s})
\end{equation}
In the table~\ref{shift} we list the measured shifts and mean
magnetization obtained by fitting the formula above to the measured
values of $G^{\op{1s}}_n(r)$ and $G^{\op{ss}}_n(r)$.  The fitting was
done by interpolating the functions by cubic splines and then
minimizing the error function~:
\begin{equation}	 
E=\sum_r
\bigl(G^{\op{AB}}_n(r)-\av{\op{A}}G^{\op{1B}}_n(r-\delta^{eff.}_\op{A})\bigr)^2
\end{equation} 
with $\delta^{eff.}_\op{A}$ and $\av{\op{A}}$ as free parameters.  As
we can see the formulas (\ref{fitform}) are very well satisfied already
for the small system sizes.

\begin{figure}[t]
\begin{center}
\psfrag{r}{$r$}
\psfrag{x}{$x$}
\includegraphics[width=10cm]{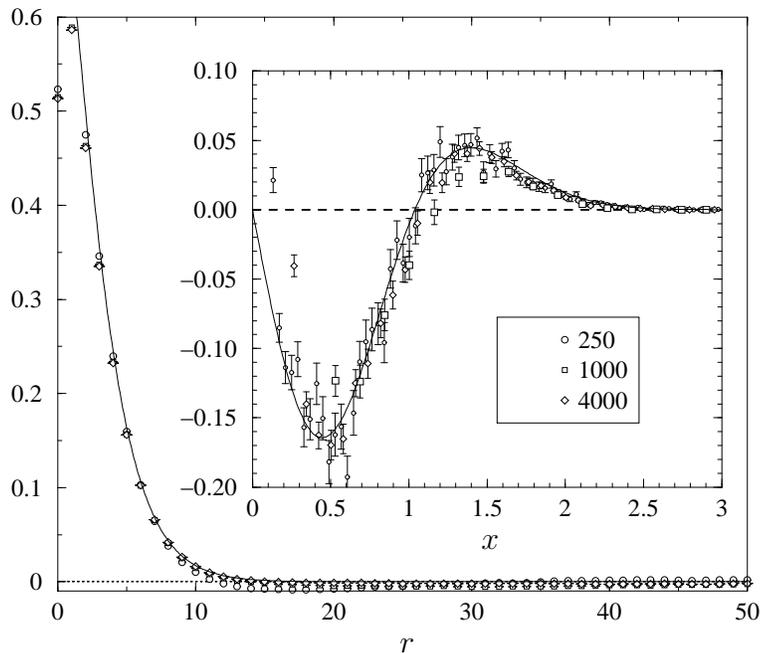}	
\end{center}
\caption{\label{ssc}Spin--spin correlation function. In the inlay the 
rescaled functions $\sqrt{n}\, G^{\op{ss}}_n(r)$ are plotted as the function
of the scaling variable $x=\frac{r}{2\sqrt{n}}$. The lines denote the theoretical predictions respectively for non--scaling and scaling parts.} 
\end{figure}

%
%
%
For the  connected spin-spin  correlator we obtain~:
\begin{equation}\label{ssnum}
X_{\op{ss}}=0.1414\qquad\text{and}\quad Y_{\op{ss}}=0.555\,.	
\end{equation}
The MC results for this correlator are plotted in the figure~\ref{ssc}. 
As expected at short distances it is dominated by
non--scaling correlations.  The  line denotes the non--scaling
part of (\ref{XY}) in the limit of $n\rightarrow\infty$. 
In the inlay we have emphasized the
scaling part by plotting the correlators scaled by $2\sqrt{n}$ as the
function of the scaling variable $x=\frac{r}{2 \sqrt{
n}}$.  The  line denotes 
the scaling part of (\ref{XY}).

\begin{figure}[t]
\begin{center} 
\psfrag{r}{$r$}
\psfrag{x}{$x$}
\includegraphics[width=10cm]{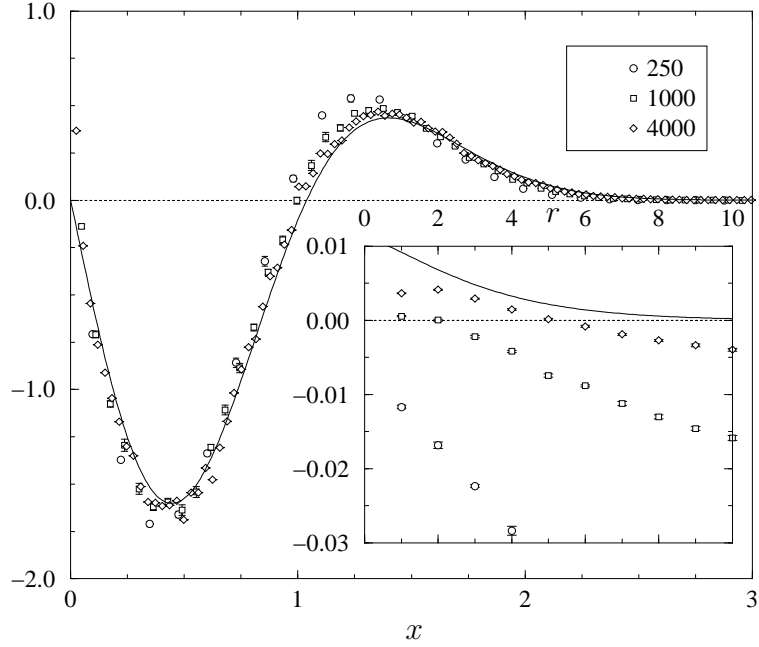}	
\end{center}
\caption{\label{qqc}``Curvature--curvature'' correlation function.
In the main part 
rescaled functions $\sqrt{n}\, G^{\op{qq}}_n(r)$ are plotted as the function
of the scaling variable $x=\frac{r}{2\sqrt{n}}$. The inlay shows unmodified 
functions. The lines denote the theoretical predictions respectively for scaling and non--scaling parts}
\end{figure}

For connected curvature--curvature correlator we obtain~:
\begin{equation}\label{qqnum}
X_{\op{qq}}=0.6861\qquad\text{and}\quad Y_{\op{qq}}=0.0070\,.	
\end{equation}
We see that in this case the non-scaling part is strongly suppressed. 
This can be seen on the figure~\ref{qqc} where we have
again shown the rescaled functions.  The non--scaling part can be seen
only by looking at small $r$ and large $n$.  In the inlay we show the
small $r$ region. 
The continuous line denotes the non--scaling part of the expression
(\ref{XY}) in the $n\rightarrow\infty$ limit.  As we increase the size
we clearly see the emerging signal of the short range correlations
which approaches the theoretical curve.

\section*{Matter--geometry interactions}

At the first sight the appearance of the long--range component
in correlation functions of non--critical fields is surprising. 
It is however clear that it is mediated through the geometry. 
To see this more explicitly we consider an example of 
 fields non-interacting with themselves~: $\ps{q,x}{q',x'}=1$
but interacting with the geometry through the term 
$\p{q}{x}$. This example can be easily solved by introducing
new weights and  operators~:
\begin{equation}	
\pp_q=\sum_x \p{x}{q}\quad\text{and}\quad
\widetilde{A}_q=\frac{\sum_x A_x \,\p{q}{x}}{\sum_x \p{q}{x}}.
\end{equation}
The correlation functions of the operators $\widetilde{A}_q$ with new
weight $\pp$ are identical to the correlation functions
of the operator $A_x$ with old weights $p$.  We see that in this case
matter operators behave as geometric operators and although matter is
non--interacting they  pick up long range correlations through
coupling to geometry.

\begin{figure}[t]
\psfrag{actual}[l][l][.9][0]{$\av{\op{s}}\;\delta_{\op{s}}$}
\psfrag{(37)}[l][l][.9][0]{$\av{\op{s}}\;\widetilde\delta_{\op{s}}$}
\psfrag{h}[l][l][1][0]{$h$}
\begin{center}
\includegraphics[width=10.0cm,bb=75 43 510 434]{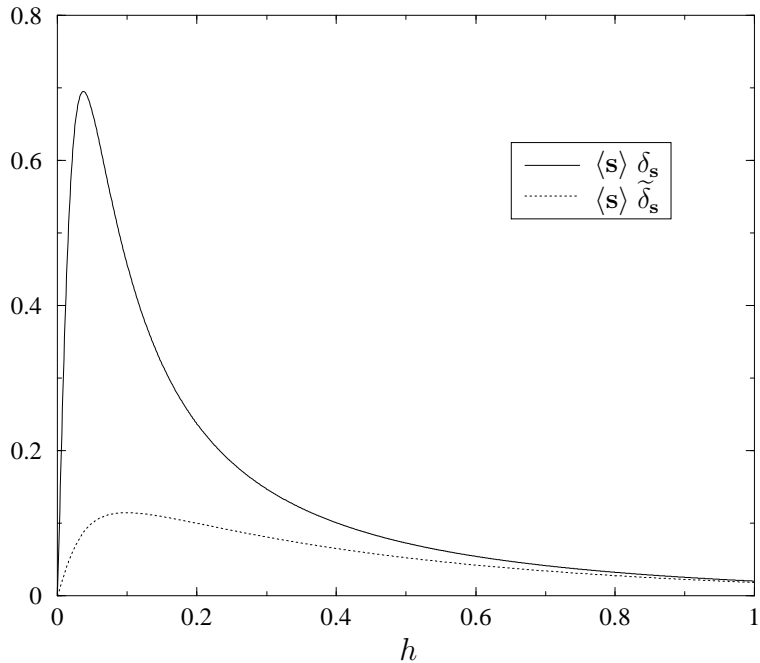}
\end{center}
\caption{\label{shiftsq}Comparison of interacting (continuous line)  and non-interacting (dotted line) spins. }
\end{figure}

To see if the effects described in previous section can be explained by
the above mechanism we assume that the spins are independent and
interact only with the geometry.  The effective coupling term
$p_{q,s}$ can be read from (\ref{avg})~:
\begin{equation}	
p_{q,s}=\frac{e^{s h}\ZGu{\mu_0}{q}{s}{q-1}\Delta Z_s(\mu_0)}{\sum_s e^{s h}F(Z_s(\mu_0))\Delta Z_s(\mu_0)}.
\end{equation}
The effective shift is then given by~:
\begin{equation}\label{shindep}	
\widetilde{\delta}^{eff}_{\op{s}}=\frac{1}{2}
\frac{\av{\op{q}}}{\av{\op{q}^2}-\av{\op{q}}^2}
\frac{\av{\op{s q}}-\av{\op{q}}\av{\op{s}}}{\av{\op{s}}}.
\end{equation}
On the figure~\ref{shiftsq}
we have plotted for comparison\footnote{$\av{\op{s}}\cdot\delta_\op{s}
\approx \delta^{eff.}_{\op{\phi}}$ where $\phi=(s+1)/2$.  This
corresponds to the definition used in \cite{Ambjorn:1998vd}.}
$\av{\op{s}}\cdot\delta_\op{s}$ and
$\av{\op{s}}\cdot\widetilde{\delta}_\op{s}$. While we see a clear
correlation, there is no quantitative agreement.  The real effect is
much more pronounced.  This is due to the non--zero spin--spin
correlation length. As a results we have an area--area like  interactions
rather then point--point like as we assumed.  When $h$ is increased
the correlation length decreases and the two values begin to agree as
expected (see figure~\ref{shiftsq}).  Probably the proper way to
proceed in this case is to use renormalization group
analysis. Blocking the spins would eventually leave us with a
non-interacting model and one could expect that the shift would be
then given by the formula (\ref{shindep}).  It is so far unclear how
to perform this blocking on the BP geometry. However a prescription
that works for Ising model on 2D random surfaces 
could provide a testing ground for this
hypothesis\cite{Thorleifsson:1996ki}.

\section*{Discussion}

We have analyzed in detail a simple model of random geometry and
derived the general structure of the correlation functions.   We
observe that in general every connected correlator contains a
weak--long--range scaling term. This term is present even in the
correlators of non--critical matter fields. The appearance of this
term is a direct consequence of the asymptotic behavior
(\ref{GABassympt}) of the disconnected correlation functions.
This behavior has a geometric origin and is picked up
 by the matter fields that effectively couple locally to the 
geometry. As we have shown in the appendix~\ref{app:symm}
this coupling is suppressed in the presence of the symmetry. 
This is intuitively clear~: When $h=0$ the probability 
of the spin being up or down must be equal independently
of the surrounding geometry.
When $h\neq0$ probability of spin being up is dependent on 
the vertex number $q$. This can be seen for example 
from the fact that in this case 
$\av{\op{sq}}\neq\av{\op{s}}\av{\op{q}}$.

The long--range scaling term can be strongly suppressed by the use
of a slightly modified definition of connected correlator (\ref{GABimp}).
This form of improved correlator postulated in \cite{Ambjorn:1998vd}
is here supported by analytical calculations.   
 
Conversely also purely geometric correlators pick up the non-scaling
behavior coming from the matter--matter interactions. In case of the
model studied here this effect is very small and  for
non--critical matter it is irrelevant in scaling limit. Nevertheless
we consider this as an interesting mechanism that could play a role in
some more complex models which would allow for the existence of the
critical matter fields.

While BP provide only a toy model of random geometry, numerical
simulations indicate that the asymptotic behavior (\ref{GABassympt})
is valid in almost every kind of random geometry ensembles.  In
particular the behavior of the Ising model in a magnetic field coupled
to the 2D simplicial quantum gravity exhibits qualitatively the same
features and the improved correlator can be successfully used
\cite{Ambjorn:1998vd}.  Similar structure is also observed for the
correlators of the local action density of the Abelian gauge fields in
the 4D simplicial quantum gravity \cite{Ambjorn:1999ix}.

It thus seems  that there exists a high degree of ``universality'' in the
structure of correlators in systems with a random geometry. Simple
models as the one described above give us so far unique
opportunity to study this structure analytically.

\section*{Acknowledgments}
The author is grateful to Z.~Burda, J.~Jurkiewicz and B.~Petersson for
many helpful comments and discussions.  This work was supported by the
Alexander von Humboldt Foundation and KBN grant 2 P03B 149 17.

\appendix

\section{\label{app:crit}Critical point} 

\newcommand{\dzo}{\Delta Z^{(1)}_{q,x}}
\newcommand{\dzt}{\Delta Z^{(2)}_{q,x}}
\newcommand{\zo}{Z_{q,x}}

\newcommand{\dzop}{\Delta Z^{(1)}_{q',x'}}
\newcommand{\dztp}{\Delta Z^{(2)}_{q',x'}}
\newcommand{\zop}{Z_{q',x'}}

Assuming the expansion~:
\begin{equation}
\ZG{\mu_0+\Dmu}{q}{x}\approx \zo(\mu_0) - \dzo(\mu_0)\sqrt{\Dmu} +
\dzt(\mu_0)\Dmu
\end{equation}
and inserting it into (\ref{sys}) 
we obtain following equations~:
\begin{align}\label{e1}	
\sum_{q',x'}&\bigl(\delta_{q,x;q',x'}-M_{q,x;q',x'}\bigr)\dzop = 0,\\
\zo =
&-\sum_{q',x'}\bigl(\delta_{q,x;q',x'}-M_{q,x;q',x'}\bigr)\dztp\nonumber\\
\label{e2}
&+\frac{1}{2}\sum_{q',x'}
\ps{q,x}{q',x'}\p{q',x'}(q'-1)(q'-2)\zop^{q'-3}(\dzop)^2.
\end{align}
Equation (\ref{e1}) states that $\dzo$ is an eigenvector
of the matrix $\op{M}(\mu_0)$ associated with the eigenvalue $\lo=1$. 
We thus have $\dzo=\mathcal{C}v^{(0)}_{q,x}$. Denoting 
by $V^{(0)}_{q,x}$ the left eigenvector of matrix $\op{M}(\mu)$ 
associated with the eigenvalue $\lo=1$  we obtain from (\ref{e2})~:
\begin{multline}	
\sum_{q,x} V^{(0)}_{q,x}\zo = \frac{1}{2}\mathcal{C}^2
e^{-\mu_0}\sum_{q,x}\sum_{q',x'}V^{(0)}_{q,x}\\
\ps{q,x}{q',x'}\p{q'}{x'}(q'-1)(q'-2)\zop^{q'-3}(v^{(0)}_{q',x'})^2.
\end{multline}
From this we can easily calculate constant $\mathcal{C}$.

\section{\label{app:tm}Transfer matrix}

Matrix $\op{\mt}$ is not symmetric and it's eigenvectors are not orthogonal.
It is more convenient to work with a symmetric matrix. To this end
we define the transfer matrix $\op{\Tt}$ by~:
\begin{equation}
\Tt_{q,x;q',x'}(\mu)=\frac{ \f{q}{x} }{ \f{q'}{x'} } M_{q,x;q'x'}(\mu)
\quad q,q'>1
\end{equation}
and zero otherwise,  where
\begin{equation}	
f_{q,x}=\sqrt{(q-1)\p{q}{x}\ZGu{\mu}{q}{x}{q-2} }.
\end{equation}
The matrix $\op{T}(\mu)$ is symmetric,  
has the same set of eigenvalues $\lk$ as
matrix $\op{\mt}(\mu)$ and its eigenvectors $\V{u}^{(k)}$ are related to the 
eigenvectors $\V{\widetilde{v}}^{(k)}$ of the matrix $\op{\mt}(\mu)$ by~:
\begin{equation}\label{u2v}
\widetilde{v}^{(k)}_{q,x}=
\begin{cases}
\displaystyle \frac{1}{\f{q}{x}}u^{(k)}_{q,x},& q>1 .   
\end{cases}
\end{equation}

In terms of matrix $\op{\Tt}$ correlation functions can be rewritten as~:
\begin{multline}	
G^{\op{AB}}_\mu(r) = \sum_{q,x}\sum_{r>1,y}\;\sum_{q',x'}\sum_{r'>1,y'}\\
\frac{\ZG{A}{q}{x}}{\f{r}{y}}M_{q,x\,;\,r,y}\;
\bigl(\Tt^{r-2}\bigr)_{r,y;r',y'}\;
(M^T)_{r',y'\,;\,q',x'}\frac{\ZG{B}{q'}{x'}}{\f{r'}{y'}}.
\end{multline}
This can be further written in terms of an orthonormal basis of eigenvectors
$\V{u}^{(k)}_{q,x}$~:
\begin{equation}\begin{split}		
G^{\op{AB}}_\mu(r)&=\sum_k \lambda_k^{r-2} \;
\sum_{q,x}\sum_{r>1,y}
\ZG{\op{A}}{q}{x}M_{q,x;r,y}\;\frac{u^{(k)}_{r,y}}{\f{r}{y}}\;\\
&\phantom{==}\kern5mm\sum_{q',x'}\sum_{r'>1,y'}
\ZG{\op{B}}{q'}{x'}M_{q',x';r',y'}\frac{u^{(k)}_{r',y'}}{\f{r'}{y'}}.
		\end{split}
\end{equation}
Using (\ref{u2v}) and the relation between eigenvectors $v$ 
of matrix $\op{M}$ and $\op{\mt}$~:
\begin{equation}\label{v2v}    
v^{(k)}_{q,x}=
\begin{cases}
\displaystyle \widetilde{v}^{(k)}_{q,x}& q>1   \\[4mm]
\displaystyle \frac{1}{\lambda_k}
\sum_{q'>1,x'}M_{1,x;q',x'}\widetilde{v}^{(k)}_{q',x'}& q=1
\end{cases}
\end{equation}
we finally obtain the expression (\ref{GABres}).

\section{\label{app:symm}Symmetry}

Let us assume that  matter field takes values in a group and that
the weights are invariant under the action of this group~: 
\begin{equation}
\ps{q,y\cdot x}{q',y\cdot x'}=\ps{q,x}{q',x'}
\quad\text{and}\quad
\p{q}{y\cdot x}=\p{q}{x}.
\end{equation}
From the above and the definition (\ref{grandpartition}) we
immediately obtain that $\ZG{\mu}{q}{x}=\ZG{\mu}{q}{1}$ and so also
$\dzo(\mu_0)=\Delta Z^{(1)}_{q,1}(\mu_0)$.  As we have shown in
appendix~\ref{app:crit} this implies $v^{(0)}_{q,y}=v^{(0)}_{q,1}$.

If  an operator $A_x$  depends only on the values of the
matter fields then~:
\begin{equation}\label{sym}
\sum_{q,x}
Z_{q,x}(A_x-\av{A})v^{(0)}_{q,x}=\sum_q
\ZG{\mu}{q}{1}p_q  v^{(0)}_{q,1} \sum_x (A_x-\av{\op{A}})
\end{equation}
But from (\ref{avg}) we have~:
\begin{equation}
\av{\op{A}}=
\frac{\sum_{q}\p{q}{1}\ZGu{\mu_0}{q}{1}{q-1}\Delta\ZG{\mu_0}{q}{1}\sum_x A_x}
{\sum_{q} \p{q}{1}\ZGu{\mu_0}{q}{1}{q-1}\Delta\ZG{\mu_0}{q}{1}\sum_x 1}
=\frac{\sum_x A_x}{\sum_x 1}
\end{equation}
which inserted into (\ref{sym}) immediately leads to~:
\begin{equation}	
\sum_{q,x}
Z_{q,x}(\op{A}-\av{\op{A}}) v^{(0)}_{q,x}\equiv 0. 
\end{equation}
This ensures that the long--range term in (\ref{GABres}) corresponding to 
the eigenvalue $\lo=1$ will be absent for any correlation function
of the type $G^{(\op{A}-\av{\op{A}})\op{B}}(r)$.

\providecommand{\href}[2]{#2}\begingroup\raggedright\endgroup

\end{document}